\documentstyle[twoside,fleqn,mymacros]{article}
\newcommand{\be}{\begin{equation}} \newcommand{\ee}{\end{equation}} \newcommand{\ba}{\begin{eqnarray}}
\newcommand{\ea}{\end{eqnarray}} \newcommand{\la}{\label}

\newcommand{\AmS}{{\protect\the\textfont2
  A\kern-.1667em\lower.5ex\hbox{M}\kern-.125emS}}

\title{On water, steam and string theory}

\author{Christof Schmidhuber
\address{Institut f\"ur Theoretische Physik, 
        Universit\"at Bern, Sidlerstr. 5, 3012 Bern, Schweiz} 
} 
\input epsf
\begin{document}

\begin{abstract}

\noindent
(Review lecture for physicists and non-physicists
as part of the requirements for ``Habilitation'' at Bern University.)

\vskip2mm
At a pressure of 220 atmospheres and a temperature of 374 ${}^0$C there is a second-order phase transition between water and steam.
Understanding it requires a key concept of elementary particle physics: the renormalization group.
Its basic ideas are explained with images from computer simulations of the lattice gas model.
It is then briefly reviewed how the renormalization group is used to compute critical coefficients for the
water-steam phase transition, in good agreement with experiment. Finally, some applications 
in particle physics and string theory are mentioned.
The appendix contains a sample of the author's results on the curious features of
renormalization group flows in theories with dynamical gravity.

\end{abstract}

\maketitle

\section{INTRODUCTION}

We all know that water can come in different phases: as a solid, a liquid or a gas.
Above $100^o$ C, water becomes steam, and below $0^o$ C it becomes ice.
We also know that the boiling and freezing temperatures of water depend on the pressure. On top of Jungfrau,
for example, water already boils at less than $90^o$ C because of the lower pressure.
The pressure dependence of the boiling and freezing temperatures of water can be plotted in a phase diagram (Fig. 1).

 \begin{figure}[htb]
 \vspace{9pt}
\vskip6cm
 \epsffile[1 1 0 0]{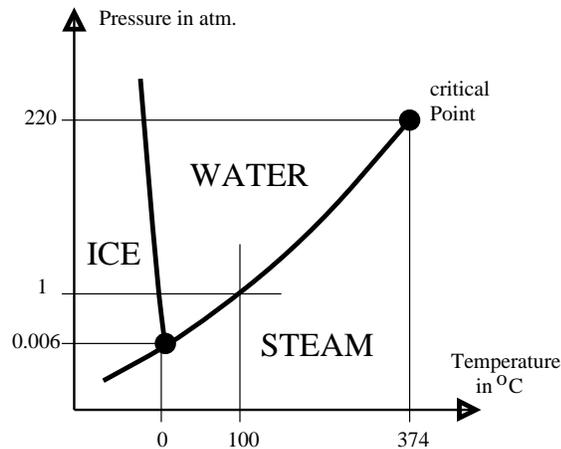}
\caption{Phase diagram of water.}
\end{figure}

Two points in this phase diagram catch one's eye. The first one is the so-called triple point at
$0^o$ C and $0.006$ atmospheres pressure.
Below this pressure the liquid phase of water disappears, and ice directly evaporates when heated up.
 {\it At} the triple point, ice, water and steam can coexist. So this is kind of interesting.
But what I would like to convince you of in my talk is that another point in this phase diagram is {\it really}
interesting, because key concepts of elementary particle physics are hidden in it. This is the so-called
critical point at $374^o$ C and 220 atmospheres pressure.

The critical point is the point at which the phase separation curve between water and steam ends.
First of all: how can the separation curve between two obviously different things such as water and steam just end?
Well, when we say that water and steam are obviously different we mean that they have drastically different
properties such as densities. Fig. 2 shows the density of water at 1 atm. pressure as a function of the temperature.
The density jumps to less than ${1\over1000}$ of its value at the point where water becomes steam.
This is called a first-order phase transition.

\begin{figure}[htb]
\vspace{9pt}
\vskip3.7cm
\epsffile[1 1 0 0]{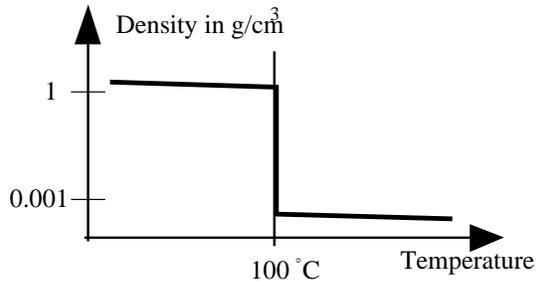}
\caption{Density of water at 1 atm. pressure.}
\end{figure}

If we now raise the pressure, we find that the densities of water and steam at boiling temperature approach each other.
The critical point is the point where both become equal, so that the distinction between water and steam seems to
disappear. Now, if we measure exactly {\it how} the densities of water and steam approach each other as we move
along the phase separation line, we find an interesting behavior: the density is not a smooth function of the
temperature, but has a singularity.

\begin{figure}[htb]
\vspace{9pt}
\vskip3cm
\epsffile[1 1 0 0]{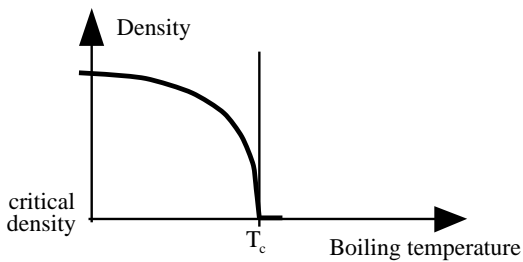}
\caption{Density of water along the phase separation curve in Fig.1.}
\end{figure}

More precisely, close to the critical point the deviation of the density $\rho$ from the critical density $\rho_c$
is proportional to a power of the deviation of the temperature $T$ from the critical temperature $T_c$:
\ba |\rho-\rho_c|\ \sim\ |T-T_c|^\beta\ .\la{anna}\ea
The so-called critical coefficient $\beta$ can be measured experimentally. One finds:
\ba \beta\ \sim\ 0.33\ \pm\ 0.01\ .\la{berta}\ea
This is called a second-order phase transition.

What I would like to do in my talk is to first give you an idea of what exactly happens at the critical point.
I will explain things with the help of images from a computer simulation of a simple toy model, the so-called lattice gas model.
As a test that this is the correct idea I will show that it indeed reproduces the power law
(\ref{anna}), and that even the coefficient $\beta$ comes out right.

Those are old ideas.
The key idea is that of the so-called renormalization group\footnote{For a review, see \cite{wilson}. 
The calculation of critical coefficients is discussed, e.g., in \cite{kadanoff,zinn}.}.
This renormalization group has
important applications in particle physics, some of which I will mention in the end.
My own results on the renormalization group in the presence of gravity and its relation to perturbative string theory
can be found in my habilitation thesis and some of them are summarized in the appendix.

{}\section{THE LATTICE GAS MODEL}

One might worry that one needs to know a lot about the structure of water molecules, hydrogen bridges, and so on,
in order to explain why $\beta$ is about $0.33$.
But surprisingly, if one measures $\beta$ for other substances with water-like phase diagrams
like Xenon or carbon dioxide, one always finds roughly $0.33$.\footnote{For a classic experiment that
measures $\beta$ in $CO_2$, see \cite{lorentzen}.} This value seems to be a universal property of
``Van-der-Waals'' gases, i.e., of gases with two basic properties: first, the water molecules have a hard core,
so that they cannot come arbitrarily close. And second, there is a weak attractive force between the molecules.

May I now introduce to you a simple toy model of such a gas, the so-called lattice gas model.
In the lattice gas model we imagine that three-dimensional space is a lattice as in Fig. 4.
Only two dimensions are drawn for simplicity. At each lattice point there is either one molecule or none.
So we equip each lattice point $i$ with an occupation number $n_i$ that is either 0 or 1.
The rule that a lattice point is occupied by at most one molecule represents the hard core of the molecules.

\begin{figure}[htb]
\vspace{9pt}
\vskip3cm
\epsffile[1 1 0 0]{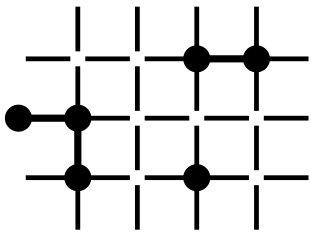}
\caption{The lattice gas model.}
\end{figure}

We define the energy of a given configuration of occupation numbers as
\ba E\ \ \ =\ \ \ -\sum_{\hbox{neighbors}\ <ij>}n_i\ \cdot\ n_j\ .\la{clara}\ea
Here, $i$ and $j$ are two neighboring lattice points. So the sum is a sum over all links.
Put differently, for each pair of neighboring molecules the energy is decreased by $-1$.
E.g., the energy of the configuration in Fig. 4 is $-3$ since there are three links that are occupied at both ends.
This definition of the energy represents the attractive force between the molecules,
because the molecules now try to stick together in order to minimize the energy.

We now imagine that the occupation numbers are subject to statistical fluctuations,
with the total number of molecules
\ba N\ =\ \sum_i\ n_i\la{diana}\ea
held fixed. Just as in a real gas the molecules can fly from one point to another, in the lattice gas they can disappear at
one point and reappear at another. 

Rather than considering a closed system with fixed particle number
it is actually more useful to consider a system that is part of a larger system with which it 
can exchange particles and with which it is in equilibrium. This is equivalent to replacing
the condition (\ref{diana}) of fixed molecule number by adding the term
\ba\mu\ \sum_i\ n_i\la{iris}\ea
to the energy $E$. $\mu$ is the so-called chemical potential whose value is controlled by the larger system.
This term punishes the smaller system for each particle it contains, thus providing a counterweight to the
attraction term
(\ref{clara}), which tries to attract as many molecules as possible.

This is the lattice gas model. You will object that the lattice gas is, at best, a rough caricature of a real gas.
But you will see that the caricature is good enough to reproduce the critical coefficient 
$\beta=0.33\pm0.01$.

{}\section{THE ISING MODEL}

To study the lattice gas model, it is useful to replace the occupation numbers $n_i$ by new variables $s_i$:
\ba s_i\ \equiv\ 2n_i\ -\ 1\ =\ \pm1\ .\la{erika}\ea
$s_i$ is $+1$ if the lattice point $i$ is occupied, and $-1$ if not.
Expressed in terms of the new variables
$s_i$, the energy (\ref{clara}+\ref{iris}) is:
\ba E\ =\ -{1\over4}\sum_{<ij>}s_is_j\ -\ B\sum_i s_i\ \la{flavia}\ea
plus an unimportant constant, where $B=1-{\mu\over2}$. 
This is the so-called Ising model, which is thus equivalent to the lattice gas model.
In the Ising model, the occupation numbers $s_i$ have the interpretation of spins that can point either up 
($s_i=+1$) or down ($s_i=-1$), while $B$ has the interpretation of an external magnetic field.
Because of the first term in the energy, neighboring spins try to point in the same direction,
while they try to point in the direction of the external magnetic field because of the second term.

What is the ground state of the Ising model - the state of lowest energy?
For $B=0$ there are two ground states: in the first one, all spins point up. In the second one, all spins point down.
Let us call them ``Plus" and ``Minus". For positive magnetic field the ground state is ``Plus".
For negative magnetic field the ground state is ``Minus".

Imagine the system is in the state ``Plus". To reverse a single spin we need
$4\cdot{1\over2}$ units of energy since the spin has 4 neighbors and for each neighboring pair we must raise the energy
from $-{1\over4}$ to $+{1\over4}$ (see Fig. 5a).
Similarly one needs ${L\over2}$ units of energy to create a ``minus bubble'' of length $L$
(Fig. 5b). One can say that bubbles have a surface tension in the sense that their energy is proportional
to the circumference of the bubble, or to the surface of the bubble in the three-dimensional model.

\begin{figure}[htb]
\vspace{9pt}
\vskip2cm
\epsffile[1 1 0 0]{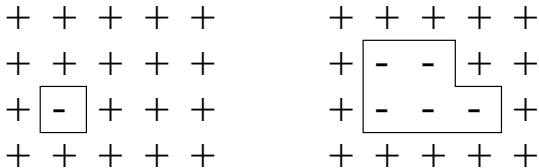}
\caption{a) A reversed spin; b) a minus bubble of length $L=10$.}
\end{figure}

We now imagine that the spins fluctuate. Let 
$\{s_i\}$ be a given configuration of all $s_i$.
As usual in Statistical Mechanics, we consider the ensemble of all possible spin configurations
$\{s_i\}$, weighted by the so-called ``Gibbs factor''
\ba \exp\{-{E(\{s_i\})\over T}\}\ ,\la{gabi}\ea
where $T$ is the temperature. The Gibbs factor can be thought of as a relative probability that is assigned to each
spin configuration. E.g., the expectation value of the spin at the point $i$ is defined as
\ba <s_i>\ =\ {1\over Z}\sum_{\{s_i\}}s_i\ \exp\{-{E(\{s_i\})\over T}\}\ .\la{hanna}\ea
$Z$ is a normalization constant,
\ba Z\ =\ \sum_{\{s_i\}}\exp\{-{E(\{s_i\})\over T}\}\ ,\la{tanja}\ea
that normalizes the probabilities such that their sum is 1. $Z$ is called the partition sum.

The Gibbs factor
(\ref{gabi}) suppresses spin configurations with high energy. More precisely, for small temperatures
all configurations are very unlikely relative to the one with lowest energy.
Fig. 6 shows images of a computer simulation of the Ising model on a lattice of size 500 times 200.\footnote{A simulation program
for the Ising model is available at  http://penguin.phy.bnl.gov/www/xtoys/xtoys.html.}
``+'' spins are black and ``--'' Spins are white.
We see two typical spin configurations at low temperature and zero magnetic field.
The system is either in the state ``Plus'' with only a few negative spins, or it is in the state ``Minus''
with only a few positive spins. Those are our two phases.
Because of their interpretation in the lattice gas model
(see (\ref{erika})) I shall call the upper phase ``Water'' and the lower phase ``steam''.

 \begin{figure}[htb]
 \vspace{9pt}
\vskip6cm
\epsffile[1 1 0 0]{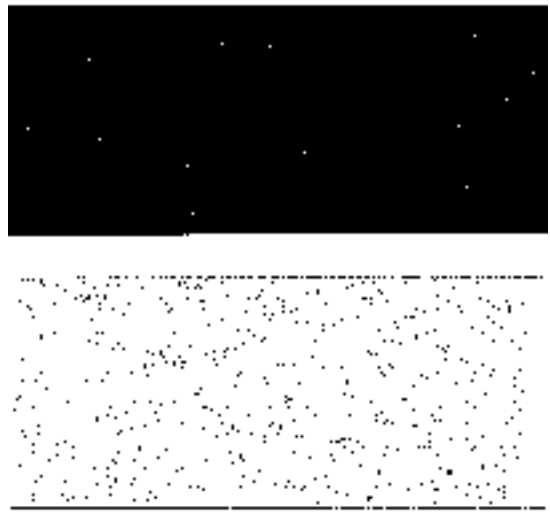}
 \caption{``Water'' (above) and ``steam'' (below) in the Ising model at low temperature.}
 \end{figure}

For very high temperatures, on the other hand, the Gibbs factor is almost one for all energies.
All spin configurations are therefore equally probable. So the typical configuration is a mix of randomly distributed spins (Fig. 7)
that is neither water nor steam. There are no different phases.

 \begin{figure}[htb]
 \vspace{9pt}
\vskip3.5cm
\epsffile[1 1 0 0]{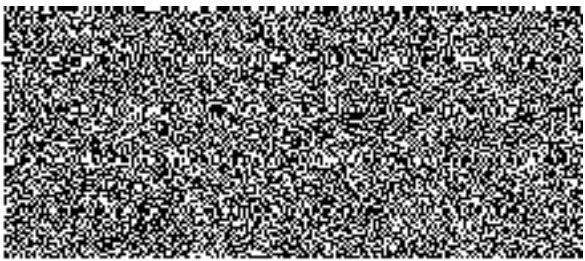}
 \caption{Ising model at very high temperature.}
 \end{figure}

We can now draw the phase diagram of the Ising model in Fig. 8. The temperature is plotted to the right and the magnetic field upwards.

 \begin{figure}[htb]
 \vspace{9pt}
\vskip5cm
\epsffile[1 1 0 0]{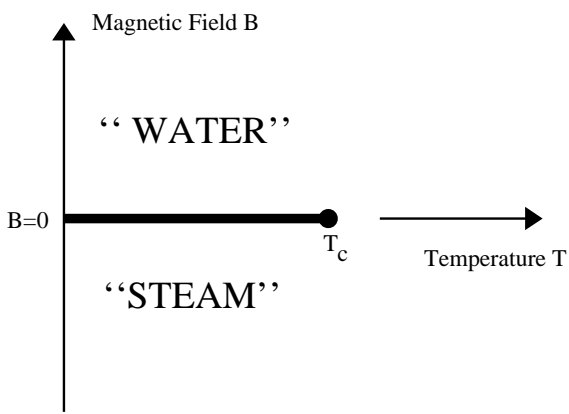}
 \caption{Phase diagram of the Ising model.}
 \end{figure}

As we have seen there is a phase separation curve at low temperature and zero magnetic field, $B=0$.
The phase transition is {\it first} order in the sense that the spin expectation value $<s_i>$ in (\ref{hanna})
- and thereby the density of molecules in the lattice gas model -- jumps as a function of $B$.
This is shown in Fig. 9. Starting from the state ``water'' at $B=0$ (upper left corner) we apply a small magnetic field $B$.
In a time-dependent simulation of the model, sooner or later bubbles appear despite of the surface tension that
are large enough to spread and turn the system into ``steam'' (lower right corner) (for $B$ constant in time).
One could say that the ``water boils''.

 \begin{figure}[htb]
 \vspace{9pt}
\vskip6.8cm
 \epsffile[1 1 0 0]{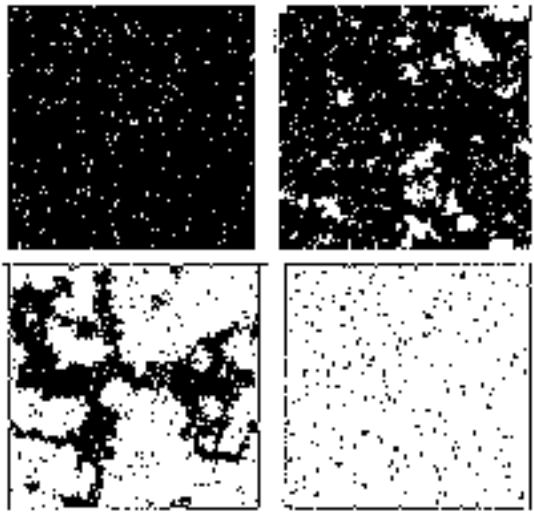}
 \caption{``Boiling'' in the Ising model.}
 \end{figure}

We have seen that there is no difference between ``water'' and ``steam'' for high temperature.
So we assume that the phase separation curve ends in a critical point at some critical temperature $T_c$.
The analogy with the phase diagram of water (Fig. 1) is clear.

{}\section{CLOSE TO THE CRITICAL POINT}\vskip2mm

Suppose the Ising model is at the phase separation curve at low temperature $T$ in the state ``water'' (e.g.,
$T={1\over2}T_c, B=0$). 
We now raise the temperature, thus approaching the critical point along the separation curve.
As $T$ grows, spin configuration with higher energy get less strongly suppressed by the Gibbs factor (\ref{gabi}).
Thus, the statistical fluctuations become more important and the typical configuration contains more and bigger
bubbles. Fig. 10 shows typical configurations for temperatures
$T=0.9\ T_c$ (above), $T=0.98\ T_c$ (middle), and $T=T_c$ (below).

 \begin{figure}[htb]
 \vspace{9pt}
\vskip11.5cm
 \epsffile[1 1 0 0]{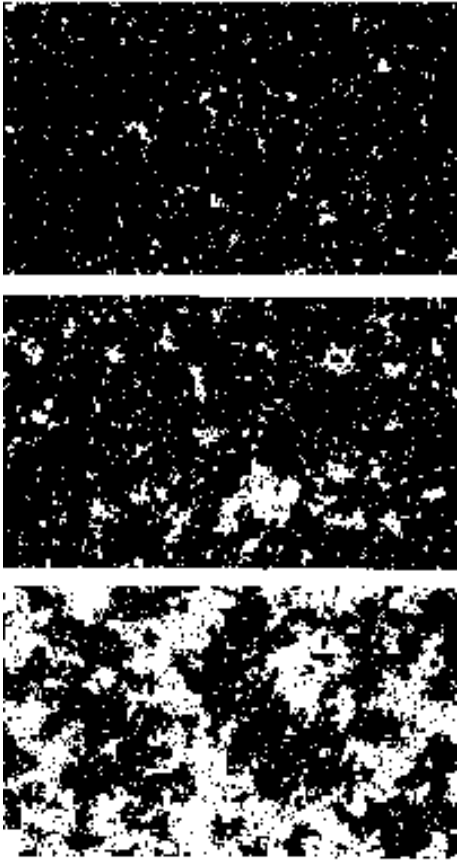}
 \caption{Approaching $T_c$ from below: $T=0.9\ T_c$ (above); $T=0.98\ T_c$ (middle); $T=T_c$ (below).}
 \end{figure}

Let us call the average size of a droplet $\xi$. You can see clearly that $\xi$ grows with the temperature.
At the critical temperature (lower picture), $\xi$ seems to diverge.
There you see black droplets inside white droplets inside black droplets and so on. You cannot really
distinguish whether the picture shows black droplets in a white sea or white droplets in a black sea.
The distinction between the phases disappears.

Let us now approach the critical temperature from above, again at zero magnetic field.
Fig. 11 shows typical configurations for temperatures $T=\infty$ (above), $T=1.08\ T_c$ (middle)
and $T=1.01\ T_c$ (below). 
At infinite temperature the spins are randomly distributed. As the temperature drops, the spins start to feel
the interaction with their neighbors through the Gibbs factor. Therefore structures form.
Let us now call $\xi$ the average size of the structures, the so-called correlation length.
You see again that $\xi$ grows as we get closer to the critical temperature.

 \begin{figure}[htb]
 \vspace{9pt}
\vskip11.5cm
 \epsffile[1 1 0 0]{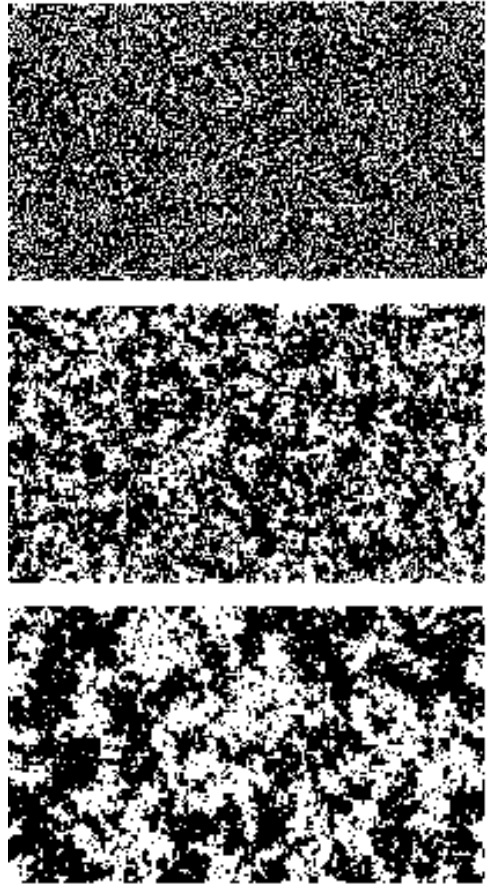}
 \caption{Approaching $T_c$ from above: $T=\infty$ (above); $T=1.08\ T_c$ (middle); $T=1.01\ T_c$ (below).}
 \end{figure}

 {\it At} the critical point it seems again as if {\it the correlation length diverges}.
Structures of all magnitudes can be recognized in a typical configuration (Fig. 12, above).
It is easy to imagine that the structures you see are only part of even larger structures that do not fit on the screen.

 \begin{figure}[htb]
 \vspace{9pt}
\vskip11.5cm
 \epsffile[1 1 0 0]{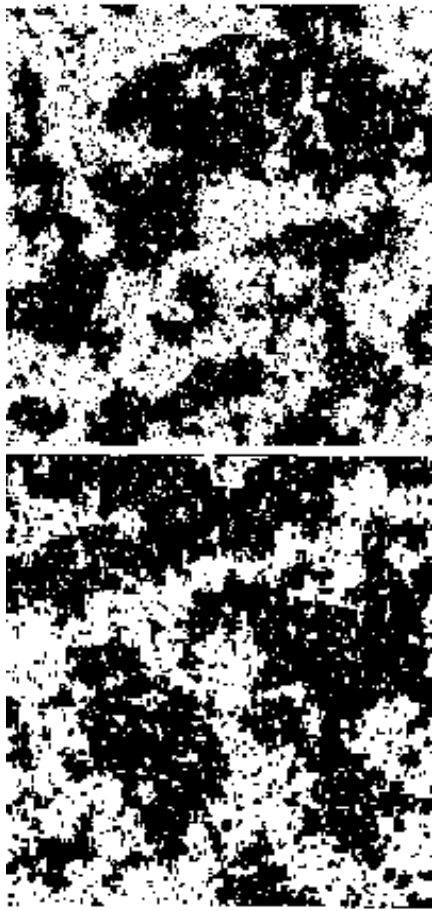}
 \caption{Ising model at the critical point (above); enlarged Ising model at the critical point (below).}
 \end{figure}

This divergence of the correlation length can also be measured experimentally in water and other substances
near the critical point. One finds that the correlation length diverges as an inverse power of the deviation of the
temperature from the critical temperature:
\ba \xi\ \sim\ {1\over |T-T_c|^{\nu}}\ ,\la{kate}\ea
with a certain coefficient $\nu$.
The relation (\ref{kate}) is similar to the relation (\ref{anna}). 
For the coefficient $\nu$ one measures:
\ba \nu\ \sim\ 0.63\pm0.01\ .\la{lisa}\ea
We will explain this behavior later, together with the behavior (\ref{anna}).

{}\section{SCALE TRANSFORMATIONS}\vskip2mm

Another property of the critical point can also be seen in Fig. 12: {\it scale invariance}.
By this, the following is meant: Fig. 12, below, shows a portion of another typical configuration
of the Ising model near the critical point, but magnified by a factor of 2. This of course makes the
picture more coarse-grained. But merely from the distribution of the sizes of the structures
you cannot determine well which of the pictures is the magnified one. 

How does the Ising model behave under scale transformations when $T$ is near, but not {\it at}
the critical temperature? Let us consider Fig. 11 again.
It seems like the lower picture could be a magnified portion of the middle picture.
So it seems as if, near the critical temperature, {\it a scale transformation is equivalent
to a change in temperature.}

Since this is a key point, let us make more precise what we mean by a scale transformation.
One can define it in two steps. The first step is to make a picture like Fig. 11, below,
more coarse-grained. E.g., one can combine $3\times3$ lattice cells into one large cell throughout the picture.
When the majority of the 9 small cells is black, the large cell is black; otherwise it is white
(Fig. 13).

 \begin{figure}[htb]
 \vspace{9pt}
 \vskip3.2cm
 \epsffile[1 1 0 0]{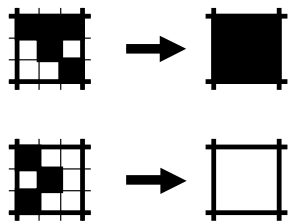}
 \caption{Increasing the lattice spacing.}
 \end{figure}

The second step consists in down-sizing the picture by a factor 3.
We call both steps together a scale transformation by a factor of 3: it is as if we regarded
the picture from 3 times as far away. The statement is that the resulting picture can be interpreted
as a portion of a typical configuration of the Ising model at a different temperature $T'$.
$T'$ is farther away from $T$ than the original temperature: the correlation length has decreased
to ${1\over3}$ of its size.

Now imagine an Ising model of infinite extent
at a temperature $T$ very close to the critical temperature. 
Let us call $T$ the ``actual'', or ``bare''
temperature. 
We consider some small portion of the model. 
We now do a scale transformation as described above by a factor $e^\tau$.
This results in an Ising model at a new temperature $T'$ that depends on $\tau$ (as opposed to $T$).
Let us call this fictitious temperature $T'$ the ``renormalized'' temperature
$T_{ren}(\tau)$. 
We have
\ba T_{ren}(\tau=0)\ =\ T\ .\ea

This dependence of the renormalized temperature $T_{ren}(\tau)$ on a change of scale by $e^\tau$
is what we call the ``renormalization group flow''
of $T_{ren}(\tau)$. The critical temperature $T_c$, at which the system is scale invariant, is
called a fixed point of this flow.

In Thermodynamics and in daily life one usually talks only about the bare temperature $T$.
On the other hand, in particle physics one usually talks about renormalized quantities
(such as charges), since those are the ones that can be observed.

Next I would like to mention how the idea of the renormalization group flow can be used to
explain the power laws (\ref{anna}) and (\ref{kate})
and to determine the coefficients $\beta$ and $\nu$.

{}\section{CRITICAL COEFFICIENTS}

Let us quickly put these statements into formulas.
We define the deviation of the temperature $T_{ren}$ from the critical temperature as $t$:
\ba t(\tau)\ =\ T_{ren}(\tau)\ -\ T_c\ .\ea
Scale invariance at the critical point means that $t=0$ is a fixed point of the flow:
\ba{d\over d\tau}t(\tau)\ =\ 0\ \ \ \ \hbox{f\"ur}\ \ \ t\ =\ 0\ .\ea
When $t$ is nonzero but small, the ``flow velocity''
${dt\over d\tau}$ can be expanded in powers of $t$:
\ba {d\over d\tau}\ t\ =\ {1\over\nu}\ t\ +\ c\ t^2\ +\ d\ t^3\ +\ ... \la{jane}\ea
with coefficients $\nu, \ c,\ d$ that need to be determined. 
The right-hand side of the equation is the so-called ``beta function'' $\beta(t)$.
The flow of $t$ is shown in Fig. 14. The arrows represent the flow velocity
$\beta(t)$. The renormalized temperature moves away from $T_c$ as the scale is increased.

 \begin{figure}[htb]
 \vspace{9pt}
 \vskip1.7cm
 \epsffile[1 1 0 0]{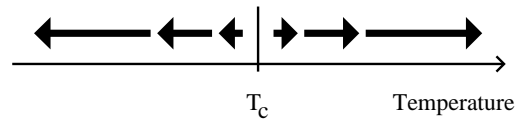}
 \caption{Flow of the renormalized temperature.}
 \end{figure}

For very small $t$, the terms $t^2,t^3,$ etc. in the beta function in (\ref{jane}) 
can be neglected.
(\ref{jane}) then becomes a linear differential equation that can easily be solved:
\ba t(\tau)\ \equiv\ |T_{ren}(\tau)-T_c|\ \sim\ |T-T_c|\ e^{{1\over\nu}\tau}\ .\la{pamela}\ea
From this, a relation between the correlation length $\xi$ and the actual temperature $T$
can be derived: as discussed in the context of Fig. 11, the correlation length shrinks
under a scale transformation by, say, a factor $e^\tau=3$ as follows:
\ba \xi\ \rightarrow\ {1\over3}\ \xi\ .\la{ulrike}\ea
From (\ref{pamela}) we read off that such a scale transformation imitates
the following change in the actual temperature:
\ba |T-T_c|\ \rightarrow\ |T-T_c|\ 3^{1\over\nu}\ .\la{veronika}\ea
(Because the scale transformation and the change in temperature then lead to the
same fictitious temperature $T_{ren}$.)
From (\ref{ulrike}) and (\ref{veronika}) we get, near the critical point, the proportionality
\ba \xi\ \sim\ {1\over |T-T_c|^{\nu}}\ ,\la{winnie}\ea
because $\xi$ then transforms as in (\ref{ulrike}) under the change of temperature (\ref{veronika}).
This is the relation (\ref{kate}), which is thereby explained by the renormalization group.
But $\nu$ still needs to be determined.

One now replaces the three-dimensional lattice again by a continuous space. One also
replaces the spin variable 
$s_i$ by a continuous field $\phi(x)$ that describes the average value of the spins in the vicinity of
the point $x$. In the lattice gas model, $\phi$ is the deviation of the density of molecules
from the critical density $\rho_c={1\over2}$:
\ba \phi(x)\ =\ \rho(x)\ -\ \rho_c\ .\la{mary}\ea
Next, one tries to describe the lattice gas model at the critical point and at large scales
by an effective field theory for the scalar field $\phi$. E.g., one replaces the partition sum
in (\ref{tanja}),
\ba \sum_{\{s_i\}}\exp\{-{E(\{s_i\})\over T}\}\ ,\ea
by a path integral
\ba\int[d\phi(x)]\ \exp\{-S[\phi(x)]\}\ ,\ea
where we formally have to do one integral for each point $x$ in space.
$S[\phi(x)]$ is a certain effective action for the field $\phi(x)$ that I will come to later.
Let me only mention now that the advantage of this continuum formulation is that it is much easier
to do computations with it than with the lattice model.
In particular, one can test whether it is really true what we have read off from Fig. 11:
is near the critical temperature and at large scales a scale transformation by the factor $e^\tau$
really equivalent to a change of the temperature $T\rightarrow T(\tau)$ ?

One finds that this is indeed the case, {\it provided that} one simultaneously allows for
a dimension $d_\phi$ of the field $\phi$. This means that $\phi$ is rescaled under scale transformations
as follows:
\ba \phi(x)\ \rightarrow\ \phi_{ren}(x)\ =\ e^{d_\phi\tau}\ \phi(x)\ .\la{nicole}\ea
Combining (\ref{pamela}), (\ref{mary}) and (\ref{nicole}), one finds similarly as in deriving
(\ref{winnie}):
\ba |\rho\ -\ \rho_c|\ \sim\ |T-T_c|^{\nu d_\phi}\ .\la{rosi}\ea
This explains the experimentally observed behavior (\ref{anna}) with $\beta=\nu d_\phi$.

It remains to compute the coefficients $\nu$ and $d_\phi$.
I cannot do this calculation in my talk and only mention that it turns out to be difficult in three
dimensions but can relatively easily be done near four dimensions with the help of the effective
field theory.
So one formally computes $\nu$ and $d_\phi$ in $4-\epsilon$ dimensions, where $\epsilon$ is small,
and then tries to set $\epsilon=1$ in the result. One finds:
\ba \nu\ &=& {1\over2}\ +\ {1\over12}\epsilon\ +\ {7\over162}\epsilon^2\ +\ ...\\
d_\phi\ &=& {1}\ -\ {1\over2}\epsilon\ +\ {1\over108}\epsilon^2\ +\ ...\ea
Therefore
\ba \beta\ =\ \nu d_\phi\ =\ {1\over2}\ -\ {1\over6}\epsilon\ +\ {1\over162}\epsilon^2\ +\ ...\ea
Those are perturbation expansions in $\epsilon$. Such expansions are useful only when they converge.
In this case we seem to be lucky: the coefficients quickly become smaller,
and the series seems to converge even in the case $\epsilon=1$, corresponding to three dimensions
(at least so it seems to this order in $\epsilon$).
If we plug in $\epsilon=1$, we get:
\ba \nu\ \sim\ 0.63\ \ \ \ ,\ \ \ \ \beta\ \sim\ 0.34\ .\la{sophie}\ \ea
This is in nice agreement with the experimental data (\ref{lisa}) and (\ref{berta}).

{}\section{UNIVERSALITY}\vskip2mm

It remains to give a reason why Xenon, carbon dioxide, water, and even the primitive lattice gas
model all produce the same critical coefficients at the critical point.
The renormalization group explains this as follows:
The previously mentioned effective action for the field $\phi(x)$ is of the form
\ba S[\phi(x)]\ =\ \int d^3x\ \{ \sum_{i=1,2,3}({\partial\phi\over\partial x_i})^2\ -\ V(\phi) \} ,\la{yvonne}\ea
where the first term measures the density fluctuations of the gas, and the second term is a potential
\ba V(\phi)\ =\ r\phi^2\ +\ \sum_{k\ge4}g_k\phi^k\ \ea
that looks as in Fig. 15 for small $\phi$ and $r<0$.
$r$ is roughly $(T-T_c)$, and the coefficients $g_k$ depend on whether one describes
$H_2O, CO_2, Xe$ or the lattice gas model (this action describes these systems only at the
phase separation line).

 \begin{figure}[htb]
 \vspace{9pt}
 \vskip4cm
 \epsffile[1 1 0 0]{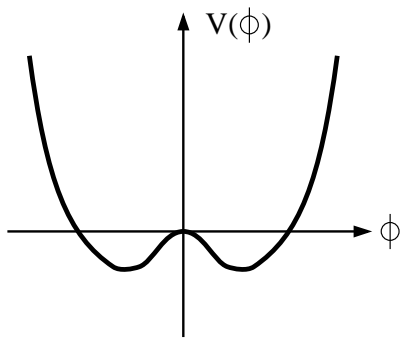}
 \caption{Effective Potential.}
 \end{figure}

One finds that all the coefficients $g_k$ flow under scale transformations. But they flow in such a way
that all points in the space of coupling constants flow at large distances towards a single fixed line,
as schematically shown in Fig. 16. In Fig. 14, {\it only} this line was drawn.

 \begin{figure}[htb]
 \vspace{9pt}
 \vskip4cm
 \epsffile[1 1 0 0]{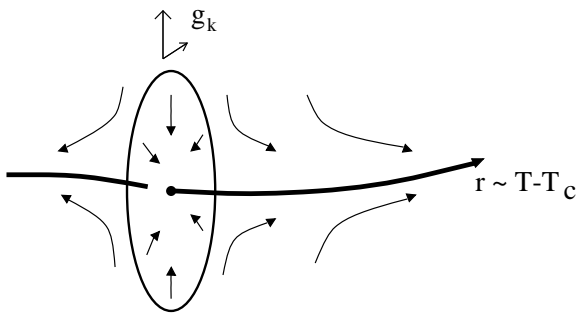}
 \caption{Flow of coupling constants.}
 \end{figure}

On this line there is a fixed point. It obviously corresponds to the critical point.
E.g., in $4-\epsilon$ dimensions it lies, to lowest order in $\epsilon$, at
\ba r\ =\ -{1\over6}\epsilon\ \ \ ,\ \ \ g_4\ =\ {1\over144}\epsilon\ \ \ ,\ \ \ea 
while the $g_k$ with $k>4$ are of order $\epsilon^2$. The only relevant parameter,
i.e. the only parameter that is observable at large distances is the parameter along the fixed line,
$(T-T_c)$. The initial (bare) values of all other parameters
$g_k$ are irrelevant in the sense that they flow to the same points at large distances.
Water, hydrogen, Xenon and the lattice gas model all have the same critical coefficients because
at large scales they are all described by the same one-parameter family of effective potentials -
those that correspond to the fixed line.

{}\section{ELEMENTARY PARTICLE THEORY}

I have told you that water and steam at the critical point are described by a three-dimensional field theory
of a scalar field. Now, there are other fields in nature. An example is the electromagnetic field
which is - among other things - responsible for the attractive force between an electron and a positron,
which is proportional to the square of the elementary charge $e$ and inversely proportional to the
square of the distance (Fig. 17).

 \begin{figure}[htb]
 \vspace{9pt}
 \vskip3cm
 \epsffile[-11 1 0 0]{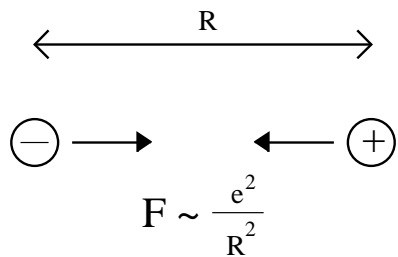}
 \caption{Force between electrons and positrons.}
 \end{figure}

Out of the elementary charge $e$, Planck's constant $\hbar$ and the speed of light $c$, a dimensionless
constant can be formed, the so-called fine structure constant $\alpha$:
\ba\alpha\ =\ {e^2\over\hbar c}\ \sim\ {1\over137}\ .\ea

The interaction of the electromagnetic field with electrons and positrons is described by a four-dimensional
field theory called quantum electrodynamics. It turns out that $\alpha$ flows under scale transformations by the
factor $e^\tau$, similarly as the temperature before. One can compute the following flow equation:
\ba {d\over d\tau}\alpha\ \sim\ -\ \alpha^2\ \ea
with a certain constant of proportionality. Here, the limit
$\tau\rightarrow\infty$ corresponds to {\it large} scales.
The corresponding flow diagram is shown qualitatively in Fig.  18.
The arrows indicate how $\alpha$ changes under a change of scale. You see that $\alpha$ decreases at large scales
and increases at small scales \cite{gellmann}.

 \begin{figure}[htb]
 \vspace{9pt}
 \vskip2cm
 \epsffile[1 1 0 0]{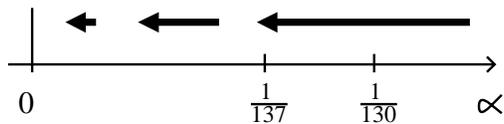}
 \caption{Flow of $\alpha$; the arrows indicate the flow as the scale is {\it increased}.}
 \end{figure}

This effect can actually be measured. When electrons and positrons are collided with high energies
so that they come close together, one finds that they scatter as if $\alpha$ was already ${1\over130}$
at scales of $10^{-17}m$.

Now, there are not only electric charges in nature but also color charges.
Quarks exist in three different ``colors'': red, green and blue.
There are also ``color forces'' between colored objects, with a corresponding interaction strength $\alpha_S$.
$\alpha_S$ is of order 1 at scales of 
order $10^{-15}m$ (this is roughly the size of a nucleus).

The interaction of the corresponding ``gluon'' fields with quarks is described by another field theory,
quantum chromodynamics. It turns out that
$\alpha_{S}$ also flows under scale transformations.
When $\alpha_{S}$ is small it obeys a differential equation of the form
\ba {d\over d\tau}\alpha_{S}\ \sim\ +\ \alpha_{S}^2\ .\ea
It is similar to the one for $\alpha$, but the constant of proportionality is different
and in particular has a {\it reversed sign.}
Therefore the arrows in the flow diagram (Fig. 19) point in the other direction.
You see that $\alpha_S$ grows at large scales and {\it decreases} at small scales \cite{gwp}.

 \begin{figure}[htb]
 \vspace{9pt}
 \vskip2cm
 \epsffile[1 1 0 0]{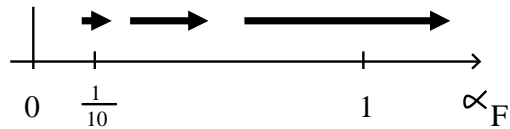}
 \caption{Flow of $\alpha_{S}$ under increasing the scale.}
 \end{figure}

This effect can also be measured. Experimentally, it seems as if $\alpha_S$ is only about ${1\over10}$
at scales of $10^{-19}m$.
Quarks seem to feel the color forces only weakly at small distances. This is called ``asymptotic freedom''.

{}\section{STRING THEORY}

There are many other interesting applications of the renormalization group in elementary particle theory,
but I want to conclude by mentioning its importance in the most ambitious branch of particle theory: in string theory.
String theory is ambitious in that it promises to unify conventional elementary particle theory
with Einstein's theory of gravity.\footnote{For an introduction to string theory, see \cite{gsw}.}

What is string theory?
I can only show two pictures in this talk. Usually it is assumed that elementary particles are pointlike,
in the sense that their classical trajectories are described by a vector
$\vec x(\sigma)$ that depends on a real parameter $\sigma$ (Fig 20).
In string theory one assumes that
$\vec x$ depends on {\it two} real parameters $\sigma_1$ and $\sigma_2$
which thereby parametrize a surface - the so-called ``world-sheet'' of the string (Fig. 21).

 \begin{figure}[htb]
 \vspace{9pt}
 \vskip4cm
 \epsffile[1 1 0 0]{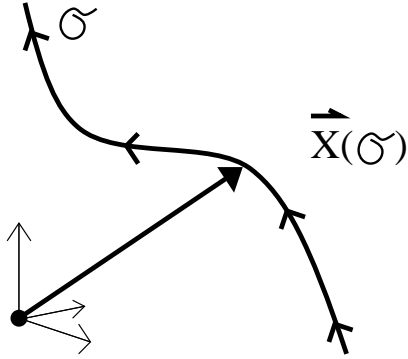}
 \caption{Trajectory of a point particle.}
 \end{figure}

 \begin{figure}[htb]
 \vspace{9pt}
 \vskip4.5cm
 \epsffile[1 1 0 0]{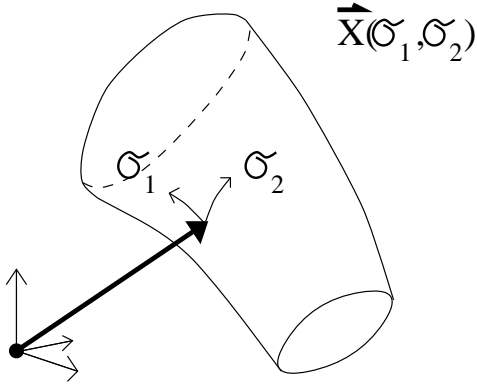}
 \caption{World-sheet of a string.}
 \end{figure}

It has proven to be very useful to reinterpret the coordinates
$\vec x(\sigma_1,\sigma_2)$ as fields that live on the world-sheet.
The motion of the string is then described by a two-dimensional field theory.
It can be shown that the corresponding action is, in the simplest case, of the form
\ba \int d^2\sigma\ \sum_{\alpha,i} ({\partial x^i\over\partial\sigma_\alpha})^2 ,\ea
where both $\alpha\in\{1,2\}$ and the index $i$ of space must be summed over.
This action is similar to (\ref{yvonne}) without potential - but the coordinates $x^i$ are now the {\it fields}.

This action can be generalized in many ways. World-sheets that live in a curved space rather than in a flat space, e.g.,
are described by replacing
\ba \sum_{\alpha,i}({\partial x^i\over\partial\sigma_\alpha})^2\ \ \rightarrow\ \ \ 
\sum_{\alpha,i,j}{\partial x^i\over\partial\sigma_\alpha}{\partial x^j\over\partial\sigma_\alpha}\ g_{ij}(\vec x)\ \ea
where $g_{ij}(\vec x)$ is the metric of the curved space.

Just like the potential in
(\ref{yvonne}) introduces a self-interaction of the field $\phi$,
$g_{ij}(\vec x)$ introduces a generally complicated interaction between the fields $x^i$.
And just like the coupling constants $g_k$ in Fig. 16 flow under scale transformations, one finds that
the metric $g_{ij}(\vec x)$ changes under scale transformations on the world-sheet.
The corresponding flow equation is quite famous \cite{friedan}:
\ba {d\over d\tau}\ g_{ij}(\vec x)\ \sim\ R_{ij}(\vec x)\ \la{zena}\ea
plus corrections, where
$R_{ij}(\vec x)$ is the curvature tensor of space at the point $\vec x$.
In particular, at a fixed point of the renormalization group flow the metric obeys Einstein's
field equations  $R_{ij} = 0$ (in vacuum) in lowest approximation.
In this sense, string theory automatically contains Einstein's theory of gravity.\footnote{That string theory 
contains gravity was found in \cite{schwarz}.}

Those two-dimensional field theories can be generalized such that, in addition to a metric, they contain
electromagnetic and other fields. The fixed point equations then become the classical
equations of motion of these interacting fields.

There are many such fixed points. Their corresponding properties directly translate into predictions for
elementary particle physics that can be compared with observation, such as the number of families of
quarks and leptons, the observable gauge group, the number of Higgs fields, etc.
An important part of (perturbative) string theory consists of finding the fixed points of the renormalization group
flow in the general two-dimensional field theory and to compute their properties.

To do this, one often uses methods that are related to those that one uses for the computation of the critical coefficients
$\beta\sim0.33$ and $\nu\sim0.63$ for the phase transition between water and steam:
the methods of the renormalization group and of ``conformal'', i.e. scale invariant field theory \cite{bpz}.
The reason for this relation between string theory and the theory of the phase transition is that
water and steam at the critical point provide a prototype of a nontrivial scale invariant field theory.

This is what I meant when I said in the beginning that the critical point is the most interesting point
in the phase diagram of water (Fig. 1).

\newpage

{}\section*{APPENDIX:\newline FLOW WITH GRAVITY}

The renormalization group comes into its own in two dimensions, where field theories often
have a rich structure of fixed points and flows between them. These flows become particularly interesting
when the theories are coupled to gravity.
Here, a sample of the author's results about renormalization group flows in two-dimensional field theories
on surfaces with fluctuating metric and topology is given.
For further results, explanations and references the author refers to his habilitation thesis,
the main part of which consists of the publications in \cite{schmi,st}.

\subsection*{A.1. Problem and motivation}
\vskip2mm

Why is it interesting to study the flow ``in the presence of gravity'' in the first place?
The motivation is two-fold.

First,
many physical systems such as QCD or the Ising model of my talk are conjectured to be
described by random-surfaces - the surfaces swept out by flux tubes or the surfaces of droplets (see e.g. \cite{pkov}).
In such cases one is often interested in predicting fixed points, critical exponents or phase diagrams.
In other words, one is interested in properties of the renormalization
group flow of the corresponding two-dimensional field theories coupled to gravity.
As is explained here for the simplest case of ``phantom surfaces'' that
do not ``feel'' self-intersections, gravity has interesting effects:
it modifies critical coefficients and phase diagrams,
and leads to curious phenomena such as oscillating flows and quantum mechanical flows.

Second, it turns out that the flow trajectories
in the presence of gravity are time--dependent classical solutions of string theory.\footnote{See, e.g., \cite{das}.}
The reason is simple. In the conformal gauge approach to quantum gravity,
the parameter $\phi$ in the conformal factor $e^\phi$ of the world-sheet metric becomes the time coordinate of the target spacetime 
\cite{polyakov,ddk}.
Thus, time translations in target space correspond to overall scale transformations,
that is, renormalization group transformations, on the world-sheet.
This suggests a rather fascinating interpretation of {\it perturbative} string theory,
that can perhaps be extended to nonperturbative string theory (which contains not only strings
but also higher-dimensional extended objects).

We thus study the following problem:
consider a renormalizable two--dimensional field theory with coupling constants
$\lambda^i$ on a surface with fixed background metric $g_{\alpha\beta}$. The coupling constants
will typically flow: the same theory that is described
at some scale $\mu$ by coupling constants $\vec\lambda$ is described
at the new scale $\mu e^{\tau}$ by new coupling constants $\vec\lambda(\tau)$.

The dependence of $\vec\lambda$ on $\tau$ is determined
by the flow equations
\ba \dot\lambda^i &=& \beta^i(\lambda^j)\la{ariane}\ea
where
\ba \beta^i &=& d^i\lambda^i\ +\ c^i_{jk}\lambda^j\lambda^k\ +\ ...\la{birgit} \ea
are the beta functions of the theory, with scaling dimensions
$d^i$ and given coefficients $c^i_{jk}$.\footnote{In this section, the limit $\tau\rightarrow\infty$
means the {\it ultraviolet}.}
How is this flow $\lambda^i(\tau)$ modified
if the theory is coupled to gravity, i.e., if the metric $g_{\alpha\beta}$ is taken to
be a dynamical variable?

{}\subsection*{A.2. ``Gravitational dressing''}
\vskip2mm

The subject of fixed points of the flow (\ref{ariane}) (i.e., of conformal field theories)
coupled to gravity was studied in \cite{polyakov}.
The ``gravitational dressing'' of the linear part of the beta functions (1.2) was derived
in light-cone gauge in \cite{kpz}, and in conformal gauge
in \cite{ddk}. The dressed dimensions $\tilde d^i$ are
\ba \tilde d^i\ =\ {Q\over\alpha}-{1\over\alpha}{\sqrt{Q^2+4d_i}}\ \ea
with certain coefficients $Q,\alpha$ that depend on the central charge $c$ of the theory at $\vec\lambda=0$.\footnote{\ \ \ 
$Q={\sqrt{{1\over3}\vert25-c\vert}}, \alpha=-{Q\over2}+{1\over2}{\sqrt{Q^2-8}}$ for $c\le1$.
For $c\ge25$ one would have $\alpha=-{Q\over2}+{1\over2}{\sqrt{Q^2+8}}$.}
One has to restrict oneself to the vicinity of fixed points
with central charge $c\le1$.

In [11a], the conformal gauge approach is extended to quadratic order in $\lambda$.
This can in particular be used to determine the
modification of the quadratic piece $c^i_{jk}$ in (\ref{birgit}) by gravity.\footnote{combining 
(2.4) and (3.1) of [11a]; see also \cite{amb}.}
One finds in the case $d^i=0$:
\begin{eqnarray}  c^i_{jk}\ \ \
\rightarrow\ \ \ \tilde c^i_{jk}\ =\ \vert{2\over Q\alpha}\vert\ c^i_{jk}\ \end{eqnarray}
in agreement with the independent light-cone gauge computation \cite{kkp}.
A modification of the cubic coefficient can also be derived but does not seem to be universal \cite{dorn}.
The supersymmetric analogs of these results have also been found \cite{gri}.

The modification of both $d^i$ and $c^i_{jk}$ in (\ref{birgit}) can be stated in an alternative way that
turns out to be more useful when one includes topology fluctuations:
gravity does not modify the beta functions in (\ref{birgit}) at all, but instead replaces the time derivative
in the flow equation (\ref{ariane}) by a second-order derivative operator \cite{st}:
\ba\dot{\vec\lambda}\ \ \ \ \rightarrow\ \ \ \ \  {\alpha^2\over4}\ \ddot{\vec\lambda}\ -\ {\alpha\over2} Q\ \dot{\vec\lambda}\ +\ 
\hbox{order($\dot\lambda^2)$}\ ,\la{christine}\ea
where $\alpha$ is negative and - to lowest order -
\ba\ Q^2 &=& {1\over3}\vert c(\vec\lambda)-25\vert\ +\ \ \hbox{order} (\dot\lambda^2)\la{daniela}\\ &\approx& {1\over3}\vert c-25\vert\ .\ea
$c(\vec\lambda)$ is the Zamolodchikov function \cite{zam}.
This gives - to the relevant order - the second-order differential equation
\ba
 {\alpha^2\over4}\ \ddot{\vec\lambda}\ -\ {\alpha\over2} Q\ \dot{\vec\lambda}\ =\ \vec\beta\ .\la{elisa}\ea 
If one now also requires $\vec\lambda$ to obey a first-order differential equation
\ba \dot{\vec\lambda}\ =\ \tilde d^i\lambda^i\ +\ \tilde c^i_{jk}\lambda^j\lambda^k\ ,\la{francis}\ea
and plugs this into (\ref{elisa}), then $\tilde d$ and $\tilde c$ are the ``modified coefficients'' mentioned above. Here one expands to linear order in $\lambda$ if $d_i\neq0$, and to quadratic order in $\lambda$
if $d_i=0$; higher orders are not universal.

(\ref{christine}) and (\ref{daniela}) are an example of the relation between
renormalization group trajectories in theories coupled to gravity and classical solutions
of string theory: they can be thought of as the string equations of motion.
(\ref{christine}) is the tachyon or graviton equation if set equal to $\vec\beta$,
while (\ref{daniela}) is the dilaton equation.

{}\subsection*{A.3. Sine-Gordon model with gravity}
\vskip2mm

One can use the conformal gauge approach to study, for example, 
the flow in the sine-Gordon model with interaction term
$m \cos px$, coupled to gravity [11a]. To the order considered, the resulting
phase diagram (Fig. 1) is in complete agreement with numerical
matrix model results \cite{grokle}:

The Kosterlitz-Thouless transition takes place in the presence of gravity,
at the same momentum $p$ as without gravity (at the boundary between
regions VI and III). In addition, there is
a new region at half the Kosterlitz-Thouless momentum (region IV). There 
appears to be a transition to negative cosmological constant, although the
interpretation has not yet been clarified.

 \begin{figure}[htb]
 \vspace{9pt}
\vskip6cm
 \epsffile[30 80 0 0]{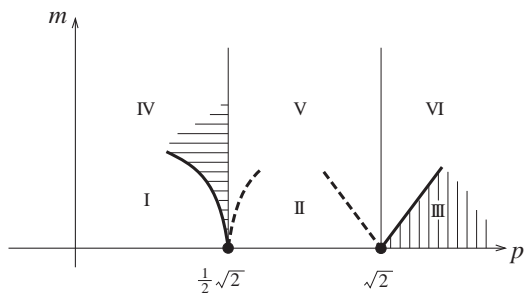}
\caption{Phase diagram of the sine-Gordon model coupled to gravity.}
\end{figure}

{}\subsection*{A.4. Topology fluctuations}
\vskip2mm

So far, the genus of the two-dimensional manifold has been assumed to be zero, corresponding to a sphere.
How is the flow modified if not only the
metric of the manifold fluctuates but also its topology? 

One then has to distinguish between effects that come from two types of pinched surfaces: pinched necks and pinched handles.
It turns out that, while metric fluctuations modify the left hand side of of the flow equation (\ref{ariane}) to (\ref{elisa}),
pinched necks modify - in addition - the right-hand side (the beta functions). Pinched handles,
on the other hand, make the flow quantum mechanical.

 \begin{figure}[htb]
 \vspace{9pt}
\vskip3cm
 \epsffile[-30 1 0 0]{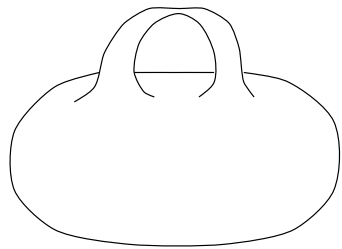}
\caption{A pinched handle.}
\end{figure}

 \begin{figure}[htb]
 \vspace{9pt}
\vskip4.5cm
 \epsffile[-30 1 0 0]{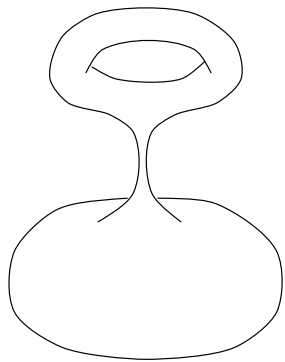}
\caption{A pinched neck.}
\end{figure}

Consider as an example a scalar matter field that lives on a circle of radius $R$ [11b].
One finds that, at genus one, random surfaces with pinched necks induce a beta function
for the radius $R$ of the circle due to the ``Fischler-Susskind mechanism'' \cite{fs}.
The radius has a fixed point at the self-dual value, $R={\sqrt 2}$ in units
where $\alpha'=2$. The deviation of $R$ from this value acquires an anomalous
dimension,
\ba d_R\ =\ -\ {1\over24}\ \kappa^2\ \ea
where
$\kappa^2$ is the topological coupling constant.

At higher genus there are further contributions to the beta functions for
both $R$ and $\kappa^2$ that can be computed in conformal gauge. A highly nontrivial consistency check yields
precise agreement with numerical (matrix model) results up to genus three.\footnote{This 
agreement was obtained in discussion with Igor Klebanov.}

Pinched handles lead to 
a curious phenomenon if the matter theory has
an isolated ``massless'' state; in the case at hand, this is the cosmological constant.
Due to a bilocal logarithmic divergence, the ``classical''
RG trajectory $\lambda^i(\tau)$ must then be replaced by a quantum mechanical
analog: to leading order, a Gaussian distribution of theories must be considered
whose square--width $\sigma^2$ also becomes a running coupling constant.
$\sigma^2$ obeys a first-order differential equation like $\lambda$ [11b]:
\ba  {\alpha^2\over4}\ {(\sigma^2)}\,\ddot{}\ -\ {\alpha\over2} Q\ {(\sigma^2)}\dot{}\ \propto\ \kappa^2\ \ea
with a beta function proportional to $\kappa^2$.

Those and all further modifications of the flow can be summarized as follows:
Topology fluctuations are accounted for by (i) ``quantizing'' the flow
and (ii) adding the higher-genus vertices
of closed string field theory to the beta functions.
This is another example of the relation between the flow with dynamical gravity and perturbative string theory.

{}\subsection*{A.5. Large central charge}
\vskip2mm

So far, we have discussed models with central charge $c\le1$ coupled to gravity. Another interesting case is that of
supersymmetric models with central charge $\hat c\ge9$ coupled to supergravity.
For simplicity, one may start by formally considering bosonic theories
with $c\ge25$ coupled to gravity (which as such make no sense) and simply ignore the tachyon instability. 
Let us also restrict ourselves to surfaces of genus zero (classical string theory).

For $c\ge25$ ($\hat c\ge9$), the conformal factor corresponds to a Minkowskian, rather than Euclidean time coordinate \cite{ddk,das}.
This has interesting consequences.
First, some minus signs get flipped, and instead of (\ref{elisa}) we now get
\ba - {\alpha^2\over4}\ \ddot{\vec\lambda}\ +\ {\alpha\over2} Q\ \dot{\vec\lambda}\ =\ \vec\beta\ .\ea

The flow towards the {\it infrared} can now be thought of as the damped motion of a particle in a potential, the generalized
Zamolodchikov function \cite{st}. 
($\alpha$ is now positive and the sign of $\tau$ is still such that $\tau\rightarrow\infty$ corresponds to the ultraviolet).
For $c\rightarrow\infty$ (infinite damping), the flow without gravity is recovered.

A novelty is that now
there can be situations where the flow towards the infrared performs damped oscillations around infrared fixed points.
In a theory without gravity, such oscillations would be forbidden
by the fact that the coupling constants at one scale uniquely determine the coupling constants at a different scale.
But when the scale is dynamical and integrated over, such an argument cannot be made. In fact, there seems to be no reason to impose
an additional condition like (\ref{francis}) for Minkowskian $\phi$.

It turns out that
for $c=25$ ($\hat c=9$), corresponding to critical string vacua,
the oscillations decay power-like rather than exponentially.

Let us mention one more aspect \cite{st}. Without gravity, the renormalization group flow of two-dimensional field theories
decreases the central charge. The generic theory that starts from an ultraviolet fixed point
with large central charge, say $c>25$ ($\hat c>9$)
will eventually - without fine-tuning of parameters - end up in the infrared at $c=0$.

It turns out that {\it with} gravity
the lowest IR fixed points that such a theory can end up at  are those with $c=25$ ($\hat c=9$) (corresponding to critical
string vacua). RG trajectories can pass through fixed points with lower $c$ but then diverge.

This is another example of the relation between
the flow with gravity and string theory:
there are two classes of renormalization group
trajectories in the presence of gravity,
corresponding to string solutions with either Minkowskian or Euclidean space-time signature.
Fixed points with $c>25$ ($\hat c>9$) lie in the former
while fixed points with $c\le1$ ($\hat c\le1$) lie in the latter class.

\newpage

{}

\noindent

\end{document}